\documentclass[conference,dvipsnames]{IEEEtran}

\IEEEoverridecommandlockouts
\usepackage{amsmath,amssymb,amsfonts}
\usepackage{graphicx}
\usepackage{xparse}
\usepackage{siunitx}
\usepackage[breaklinks,colorlinks]{hyperref}
\usepackage[capitalize]{cleveref}

\usepackage[style=ieee, backend=bibtex, defernumbers=true]{biblatex}
\usepackage{xcolor} 

\usepackage[draft]{pdfcomment} 
\usepackage[hyperfirst=false,acronym,sort=none,shortcuts,nopostdot,style=super,nonumberlist,toc,nogroupskip]{glossaries}

\glsdisablehyper 

\newcommand{\accolor}[1]{\textcolor{Sepia}{#1}}

\newacronymstyle{myacro}
{%
  \GlsUseAcrEntryDispStyle{long-short}%
}%
{%
  \GlsUseAcrStyleDefs{long-short}%
}

\setacronymstyle{myacro}

\newcommand*{\tip}[1]{
    \ifglsused{#1}{
      {\pdftooltip{\accolor{\glsentryshort{#1}}}{\glsentrydesc{#1}}}%
    }{%
      \gls{#1}
    }%
}%

\newcommand*{\tips}[1]{
    \ifglsused{#1}{
      {\pdftooltip{\accolor{\glsentryshortpl{#1}}}{\glsentrydescplural{#1}}}%
    }{%
      \glspl{#1}
    }%
}%

\newacronym{DVS}{DVS}{Dynamic Vision Sensor}
\newacronym{PSD}{PSD}{Power Spectral Density}
\newacronym{TC}{TC}{Temporal Contrast}
\newacronym{SNR}{SNR}{Signal-to-Noise Ratio}
\newacronym{RMS}{RMS}{root mean square}
\newacronym{SF}{SF}{Source-Follower buffer}
\newacronym{theta_on}{$\Theta_{\text{ON}}$}{ON threshold}
\newacronym{theta_off}{$\Theta_{\text{OFF}}$}{OFF threshold}

\newacronym[description={False Negative Rate; signal that is incorrectly classified as noise}]{fnr}{FNR}{False Negative Rate}
\newacronym[description={False Negative; signal event that is incorrectly classified as noise event}]{fn}{FN}{False Negative}
\newacronym[description={False Positive Rate; noise that is incorrectly classified as signal}]{fpr}{FPR}{False Positive Rate}
\newacronym[description={False Positive; noise event that is incorrectly classified as signal event}]{fp}{FP}{False Positive}
\newacronym[description={Guided Event Filtering: Joint Filtering of Intensity Images and Neuromorphic Events}]{gef}{GEF}{Guided Event Flow}
\newacronym[description={Inter Spike Interval (nomenclature from neuroscience)}]{isi}{ISI}{Inter Spike Interval}
\newacronym[description={MAC (Multiply-Accumulate) is the basic operation of signal processing and artificial neural networks. One MAC is 2 Op.}]{mac}{MAC}{Multiply-Accumulate}
\newacronym[description={Multipurpose block random access memory module in FPGA}]{bram}{BRAM}{Block RAM}
\newacronym[description={Register Transfer Logic intermediate form, consisting of combinational and synchronous register logic cells}]{rtl}{RTL}{Register Transfer Logic}
\newacronym[description={Single Threshold Metric; a measure of the ROC TPR/FPR tradeoff at one discrimination threshold}]{stm}{STM}{Single Threshold Metric}
\newacronym[description={Surface of Active Event; image of latest event timestamps, same as Timestamp Image}]{sae}{SAE}{Surface of Active Events}
\newacronym[description={System on Chip; FPGA with embedded programmable processor}]{soc}{SoC}{System on Chip}
\newacronym[description={Time Surface; image of age of events relative to a particular event}]{ts}{TS}{Time Surface}
\newacronym[description={Timestamp Image; 2D image of latest event timstamps, similar to Surface of Active Events}]{ti}{TI}{Timestamp Image}
\newacronym[description={Timestamp+Polarity Image; 2D image of latest event timstamps and +/- brightness change polarites}]{tpi}{TPI}{Timestamp+Polarity Image}
\newacronym[description={True Negative Rate; noise that is correctly classified as noise}]{tnr}{TNR}{True Negative Rate}
\newacronym[description={True Negative; noise that is correctly classified as noise}]{tn}{TN}{True Negative}
\newacronym[description={True Positive Rate; signal that is correctly classified as signal}]{tpr}{TPR}{True Positive Rate}
\newacronym[description={True Positive; signal event that is correctly classified as signal}]{tp}{TP}{True Positive}
\newacronym[longplural={Convolutional Neural Networks}]{cnn}{CNN}{Convolutional Neural Network}
\newacronym[longplural={First In First Out memories}]{fifo}{FIFO}{First In First Out memory}
\newacronym{adc}{ADC}{Analog to Digital Converter}
\newacronym{aer}{AER}{Address Event Protocol}
\newacronym{aps}{APS}{Active Pixel Sensor}
\newacronym{asic}{ASIC}{Application Specific Integrated Circuit}
\newacronym{auc}{AUC}{Area Under the Curve}
\newacronym{baf}{BAF}{Background Activity Filter}
\newacronym{ba}{BA}{Background Activity}
\newacronym{bmof}{BMOF}{Block Matching Optical Flow}
\newacronym{bm}{BM}{Block Matching}
\newacronym{bp}{BP}{Back Propagation}
\newacronym{cfa}{CFA}{Color Filter Array}
\newacronym{cf}{CF}{Complementary Filter}
\newacronym{cg}{CG}{Convolutional Gated Recurrent Unit Network}
\newacronym{cis}{CIS}{CMOS Image Sensor}
\newacronym{cmae}{CMAE}{Cross-Modality Attention Enhancement}
\newacronym{contrastmaximization}{CM}{Contrast Maximization}
\newacronym{cots}{COTS}{Commodity Off-The-Shelf}
\newacronym{cpu}{CPU}{Central Processing Unit}
\newacronym{cv}{CV}{Computer Vision}
\newacronym{davis}{DAVIS}{Dynamic and Active pixel Vision Sensor}
\newacronym{dba}{DBA}{Dynamic Background Activity noise filtering algorithm}
\newacronym{dnn}{DNN}{Deep Neural Network}
\newacronym{dof}{DOF}{Degree of Freedom}
\newacronym{dolp}{DoLP}{Degree of Linear Polarization}
\newacronym{dram}{DRAM}{Dynamic RAM}
\newacronym{drcn}{DRCN}{Deep Recurrent Convolutional Network}
\newacronym{dr}{DR}{Dynamic Range}
\newacronym{dsp}{DSP}{Digital Signal Processing unit}
\newacronym{dvs}{DVS}{Dynamic Vision Sensor}
\newacronym{dwf}{DWF}{Double Window Filter}
\newacronym{e2pd}{E2PD}{Events to Polarization Dataset}
\newacronym{e2p}{E2P}{Events to Polarization}
\newacronym{edflow}{EDFLOW}{Event-driven Optical Flow}
\newacronym{edncnn}{EDnCNN}{Event Denoising CNN}
\newacronym{edp}{EDP}{Event Denoising Precision}
\newacronym{efast}{EFAST}{Event-Based time surface FAST}
\newacronym{epm}{EPM}{Event Probability Mask}
\newacronym{fast}{FAST}{Features from Accelerated Segment Test}
\newacronym{feast}{FEAST}{Feature Extraction with Adaptive Selection Thresholds }
\newacronym{flipflop}{FF}{Flip-Flop}
\newacronym{fom}{FOM}{Figure of Merit}
\newacronym{fpga}{FPGA}{Field Programmable Gate Array}
\newacronym{fpn}{FPN}{Fixed Pattern Noise}
\newacronym{fps}{FPS}{Frames Per Second}
\newacronym{fsae}{FSAE}{Filtered Surface of Active Events}
\newacronym{fwf}{FWF}{Fixed Window Filter}
\newacronym{gpu}{GPU}{Graphics Processing Unit}
\newacronym{gt}{GT}{Ground Truth}
\newacronym{hdl}{HDL}{Hardware Description Language}
\newacronym{hdr}{HDR}{high dynamic range}
\newacronym{hls}{HLS}{High Level Synthesis}
\newacronym{icm}{ICM}{Iterated Conditional Modes}
\newacronym{id}{ID}{Index Decay}
\newacronym{iir}{IIR}{Infinite Impulse Response}
\newacronym{imu}{IMU}{Inertial Measurement Unit}
\newacronym{inceptiveevent}{IE}{Inceptive Event}
\newacronym{iot}{IoT}{Internet of Things}
\newacronym{ip}{IP}{Intellectual Property}
\newacronym{its}{ITS}{Invariant Time Surface}
\newacronym{knn}{KNN}{$K$-Nearest-Neighbor clustering}
\newacronym{li}{LI}{Leaky Integrator}
\newacronym{lk}{LK}{Lucas-Kanade}
\newacronym{lpips}{LPIPS}{Learned Perceptual Image Patch Similarity}
\newacronym{lut}{LUT}{LookUp Table}
\newacronym{mlpf}{MLPF}{MultiLayer Perceptron denoising Filter}
\newacronym{mlp}{MLP}{Multilayer Perceptron}
\newacronym{ml}{ML}{Machine Learning}
\newacronym{mpeg}{MPEG}{Motion Picture Experts Group}
\newacronym{mse}{MSE}{Mean Squared Error}
\newacronym{na}{NA}{Numerical Aperture}
\newacronym{nnb}{NNb}{Nearest Neighbor}
\newacronym{of}{OF}{Optical Flow}
\newacronym{onf}{ONF}{Order(N) Filter}
\newacronym{pcb}{PCB}{Printed Circuit Board}
\newacronym{pdavis}{PDAVIS}{Polarization Dynamic and Active pixel VIsion Sensor}
\newacronym{pd}{PD}{photodiode}
\newacronym{pe}{PE}{Processing Element}
\newacronym{pfa}{PFA}{Polarization Filter Array}
\newacronym{pl}{PL}{programmable Logic}
\newacronym{por}{POR}{Positive Output Ratio}
\newacronym{prm}{PRM}{Pixel Rendering Module}
\newacronym{ps}{PS}{Processing System}
\newacronym{pugm}{PUGM}{Probabilistic Undirected Graph Model}
\newacronym{qwp}{QWP}{Quarter Wave Plate}
\newacronym{ram}{RAM}{Random Access Memory}
\newacronym{ransac}{RANSAC}{Random Sample and Consensus}
\newacronym{ratp}{RATP}{Recursive Adaptive Temporal Pooling}
\newacronym{rb}{RB}{Residual Block}
\newacronym{relu}{ReLU}{Rectified Linear Unit}
\newacronym{roc}{ROC}{Receiver Operating Characteristic}
\newacronym{roi}{ROI}{Region of Interest}
\newacronym{rpmd}{RPMD}{Relative Plausibility Measure of Denoising}
\newacronym{rpm}{RPM}{Revolutions per Minute}
\newacronym{rppp}{RPPP}{Rich Polarization Pattern Perception}
\newacronym{sad}{SAD}{Sum of Absolute Differences}
\newacronym{sm}{SM}{Supplementary Material}
\newacronym{snr}{SNR}{Signal to Noise Ratio}
\newacronym{soa}{SOA}{state of the art}
\newacronym{sram}{SRAM}{Static RAM}
\newacronym{stcf}{STCF}{SpatioTemporal Correlation Filter}
\newacronym{tda}{TDA}{Time Decay Adapted}
\newacronym{td}{TD}{Time Decay}
\newacronym{timsl}{TS}{time slice}
\newacronym{usb}{USB}{Universal Serial Bus}
\newacronym{vga}{VGA}{Video Graphics Adaptor}
\newacronym{vhdl}{VHDL}{Very High-Speed Integrated Circuit Hardware Description Language}
\newacronym{zoh}{ZOH}{Zero-Order Hold}

\newcommand{\Mfb}{\text{M}_\text{fb}}

\newcommand{\Mn}{\text{M}_\text{n}}

\newcommand{\ipr}{I_\text{pr}}

\newcommand{\ipd}{I_\text{pd}}
\newcommand{\vpr}{V_\text{pr}}
\newcommand{\isf}{I_\text{sf}}
\newcommand{\vsf}{V_\text{sf}}

\newcommand{\trefr}{\Delta_\text{refr}}

\begin{document}

\title{Optimal biasing and physical \\ limits of DVS event noise}

\author{\IEEEauthorblockN{Rui Graca, Brian McReynolds, Tobi Delbruck}
\IEEEauthorblockA{\textit{Sensors Group, Inst. of Neuroinformatics, UZH-ETH Zurich, 
Zurich, Switzerland} \\
rpgraca,bmac,tobi@ini.uzh.ch, \url{https://sensors.ini.uzh.ch}}
}

\maketitle
\thispagestyle{plain}
\pagestyle{plain}


\begin{abstract}
Under dim lighting conditions, the output of Dynamic Vision Sensor (DVS) event cameras is strongly affected by noise. Photon and electron shot-noise cause a high rate of non-informative events that reduce Signal to Noise ratio. DVS noise performance depends not only on the scene illumination, but also on the user-controllable biasing of the camera. In this paper, we explore the physical limits of DVS noise, showing that the DVS photoreceptor is limited to a theoretical minimum of 2x photon shot noise, and we discuss how biasing the DVS with high photoreceptor bias and adequate source-follower bias approaches optimal noise performance. We support our conclusions with pixel-level measurements of a DAVIS346 and analysis of a theoretical pixel model.
\end{abstract}

\section{Introduction}
\label{sec:intro}
\begin{figure*}[tb]
    \centering
    \includegraphics[width=\textwidth]{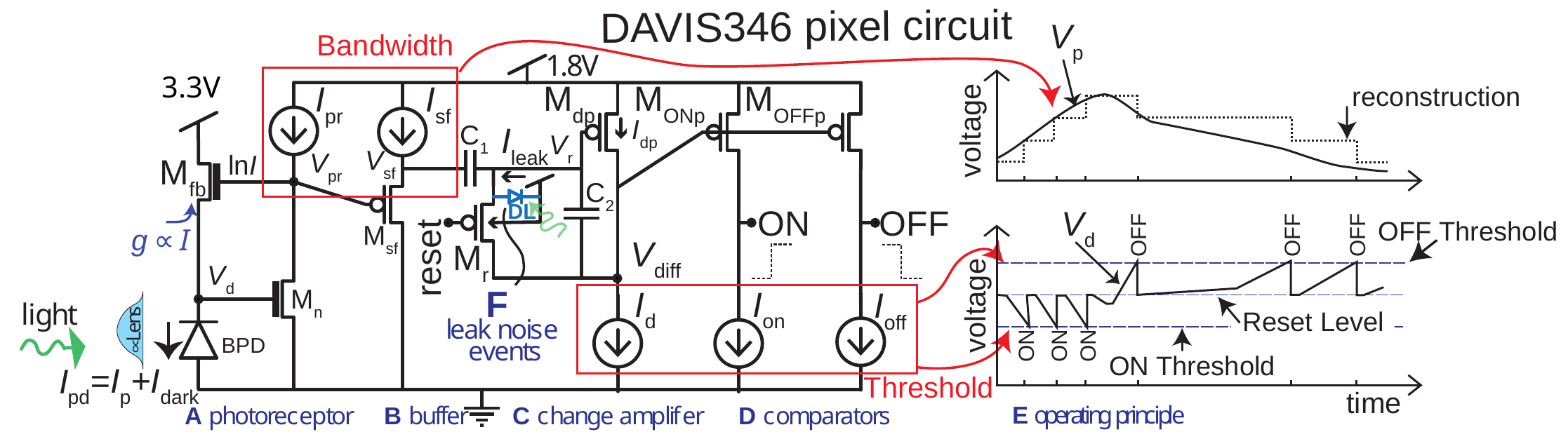}
    \caption{Typical DVS pixel circuit~\cite{taverni2018frontandback}. The active logarithmic photoreceptor (\textbf{A}) is buffered by a source-follower (\textbf{B}), which drives a cap-feedback change amplifier (\textbf{C}), which is reset on each event by a low-going \textit{reset} pulse.
    A finite refractory period holds the change amplifier in reset for the refractory period $\trefr$. Comparators (\textbf{D}) detect ON and OFF events as seen in \textbf{E}. Periodic leak events result from junction and parasitic photocurrent $I_\text{leak}$ in diode \textsf{DL} (\textbf{F}). 
    }
    \label{fig:davis_circuit}
\end{figure*}

The \tip{DVS}~\cite{lichtsteiner2008latencyasynchronous,suh2020dynamicvision,finateu2020backilluminated,brandli2014latencyglobal} is a neuromorphic event-based vision sensor, which consists of an array of asynchronously operating pixels as the one in \cref{fig:davis_circuit}~\cite{brandli2014latencyglobal}. Each pixel independently encodes instantaneous changes in its input light into an asynchronous steam of ON and OFF events. More specifically, a pixel outputs an ON event when the relative \tip{TC}~\cite{lichtsteiner2008latencyasynchronous} of light intensity at its input increases by a user defined ON threshold since the last event, or an OFF event when the relative \tip{TC} increases by a user defined OFF threshold since the last event. When its input is static, a \tip{DVS} pixel ideally outputs no event. 
More extensive description of the \tip{DVS} pixel operation can be found in ~\cite{lichtsteiner2008latencyasynchronous,hu2021v2e,graca2023shininglight}.

\begin{figure*}
    \centering
    \includegraphics[width=\textwidth]{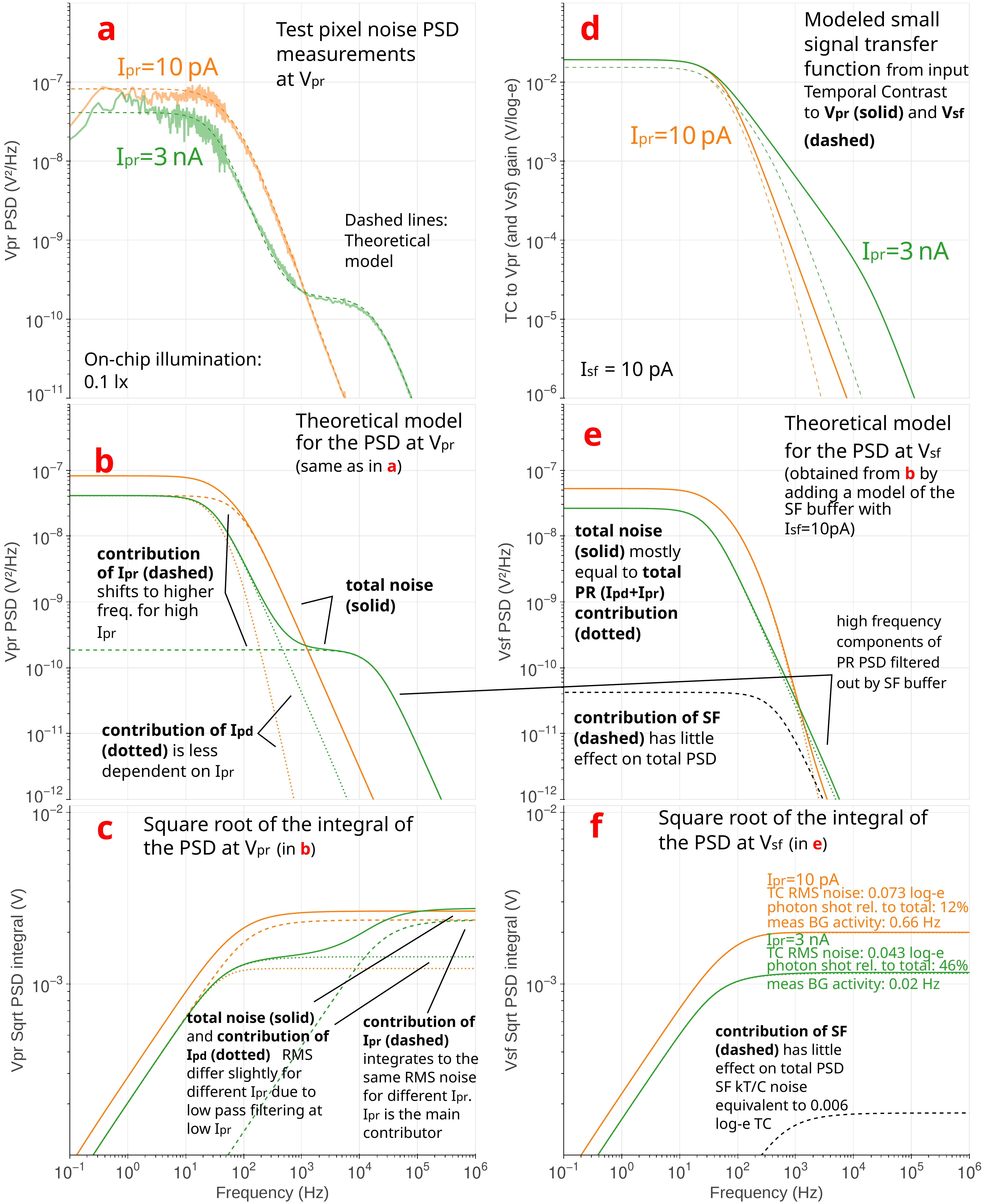}
    \caption{(\textbf{a}) shows noise \tips{PSD} measured from a DAVIS346 test pixel for two different $\ipr$ biases for an on-chip illuminance of \SI{0.1}{\lux} and the \tips{PSD} estimated by a theoretical model for the same conditions. (\textbf{b}) shows the estimated contributions of $\ipr$ and $\ipd$ to the total PSD for the same model in the same conditions, and (\textbf{c}) shows the square root of the integral of the curves in (\textbf{b}). The final value of these curves is the respective contribution to the \tip{RMS} noise voltage at $\vpr$. (\textbf{e}) and (\textbf{f}) show the same quantities as (\textbf{b}) and (\textbf{c}), but relative to $\vsf$. (\textbf{d}) shows the estimated signal transfer function from \tip{TC} (in log-e units) to $\vpr$ and $\vsf$. }
    \label{fig:psd}
\end{figure*}

Characteristics of the \tip{DVS} such as sparse data encoding and low latency make it a good candidate for scientific applications such as space situational awareness and wide-field voltage and calcium imaging. The adequacy of the \tip{DVS} for some applications is potentially limited by a too high rate of parasitic \tip{ba}. \tip{ba} consists of events that do not encode changes in the input. These events are undesirable because they decrease the \tip{SNR} and increase data volume~\cite{hu2021v2e,guo2023lowcost}. The \tip{ba} of the \tip{DVS} pixel strongly depends on both light intensity and camera biasing~\cite{finateu2020backilluminated,graca2021unravelingtheparadox,nozaki2017temperatureandparasitic,graca2023shininglight,mcreynolds2023exploitingalternating}. It is predominantly caused by photon and electron shot noise in dark settings~\cite{graca2021unravelingtheparadox}, and by leakage in the reset transistor (\cref{fig:davis_circuit}\textbf{F}) in brighter settings~\cite{nozaki2017temperatureandparasitic}. 

A good understanding of the phenomena resulting in \tip{ba} is important for improving camera models that can aid pixel design, optimization of the camera utilization, or learning algorithms~\cite{hu2021v2e,graca2023shininglight,mcreynolds2022experimentalmethods,delbruck2021feedbackcontrol}. 

In~\cite{graca2021unravelingtheparadox}, noise power at the output of the photoreceptor ($\vpr$ in \cref{fig:davis_circuit}) and noise event rate are explored as a function of illumination and photoreceptor bias $\ipr$. There, we observe that both noise power and event rate are lower for lower $\text{I}_{\text{pr}}$. This occurs because the bandwidth is lower for lower $\text{I}_{\text{pr}}$.

These observations suggest that using a small $\text{I}_{\text{pr}}$ to limit bandwidth reduces noise, and this assumption has been used as an optimization rule for bias control~\cite{delbruck2021feedbackcontrol}. In this paper, we go a step further into understanding the optimal conditions and biasing of the \tip{DVS} pixel, and show that in fact the opposite is generally true -- even though strongly reducing $\ipr$ leads to a decrease in noise events, noise performance is more optimal for high $\ipr$. We show that the \tip{DVS} photoreceptor topology is bounded with a theoretical minimum of 2x photon shot noise, and we discuss bias optimization regarding bandwidth and its implications on noise and signal. In this paper, we focus on the biasing of the photoreceptor (\cref{fig:davis_circuit}\textbf{A}) by $\ipr$ and the \tip{SF} (\cref{fig:davis_circuit}\textbf{B}) by $\isf$. A more general discussion about bias optimization is presented in~\cite{graca2023shininglight}, and considerations about threshold and refractory biases are discussed in~\cite{mcreynolds2023exploitingalternating}.

\section{Optimal Photoreceptor biasing}
\label{sec:optimal_biasing}

\subsection{PSD Measurements and modeling}
\cref{fig:psd}\textbf{a} shows the noise \tip{PSD} measured at $\vpr$ of a test pixel isolated from a DAVIS346 array under an on-chip illuminance of \SI{0.1}{\lux} for two different $\ipr$ settings: one high (\SI{3}{\nano\ampere}) and one low (\SI{10}{\pico\ampere}). The dashed lines in the figure show the \tip{PSD} predicted by a theoretical physically-realistic model operating under the same conditions. The theoretical model was obtained by circuit analysis considering the sources of shot noise in the photoreceptor and applying the transfer function that relates them to $\vpr$. The parameters for the model were then estimated and fitted based on SPICE simulation and pixel measurements.

Since the theoretical model generally matches both measured and simulated data, we utilize it to further infer about the noise contribution of each noise source to the total output noise. In \cref{fig:psd}\textbf{b}, we see how the contribution of the photocurrent $\ipd$ (depicted by the dotted lines and consisting of photon shot noise at the photodiode and electron shot noise added by $\Mfb$) and the contribution of $\ipr$ (depicted by the dashed lines, and consisting of noise introduced by $\Mn$ and the transistor implementing $\ipr$) add up to the total \tip{PSD}. Here, we observe that the level of the contribution of $\ipd$ is independent of $\ipr$, but its bandwidth may depend on $\ipr$ -- for a bias of \SI{10}{\pico\ampere}, $\ipr$ is right at the edge of starting to filter out the $\ipd$ contribution. That is, this contribution would be significantly reduced for lower $\ipr$, and would become constant for higher $\ipr$ (as happens for $\ipr$ of \SI{3}{\nano\ampere}. On the other hand, the contribution of $\ipr$ moves to higher frequencies when $\ipr$ increases. 

\cref{fig:psd}\textbf{c} shows the square root of the integral of the different components of the \tip{PSD} in \cref{fig:psd}\textbf{b}. The final value of the square root of the integral is the \tip{RMS} voltage noise contribution of its respective source. We see that the contribution of $\ipr$ converges to a value independent of $\ipr$ - the contribution is only shifted to higher frequencies. The contribution of $\ipd$ is lower for lower $\ipr$, which happens due to filtering by $\ipr$ \cite{graca2021unravelingtheparadox}. For higher values of $\ipr$, filtering would stop occurring and the contribution of $\ipd$ converges to the constant value observed at $\ipr$ of \SI{3}{\nano\ampere}.

Figs. \ref{fig:psd}\textbf{e} and \ref{fig:psd}\textbf{f} show the modeled \tips{PSD} and the square root of their integrals for the contributors at the output of the \tip{SF}, $\vsf$. The \tips{PSD} were obtained by filtering the ones at $\vpr$ using a model of the \tip{SF} estimated by circuit inspection and simulation. Also, the noise contribution of $\isf$ is added in the dashed line. However, its value is much smaller than the contribution of the photoreceptor (the summation of the contributions of $\ipd$ and $\ipr$). 

\cref{fig:psd}\textbf{d} shows the modeled signal transfer function from logarithmic changes in light intensity to voltages at $\vpr$ and $\vsf$. As described in~\cite{graca2021unravelingtheparadox}, it can be approximately modeled as a second order system with one pole dependent on $\ipd$ and the other dependent on $\ipr$. At $\ipr$ of \SI{3}{\nano\ampere}, the pole controlled by $\ipd$ is clearly dominant, while for $\ipr$ of \SI{10}{\pico\ampere} the two poles lie very close to each other. The \tip{SF} add another pole, which for the bias used is close to the dominant of the photoreceptor.

\begin{figure*}[t]
\centering
\includegraphics[width=\textwidth]{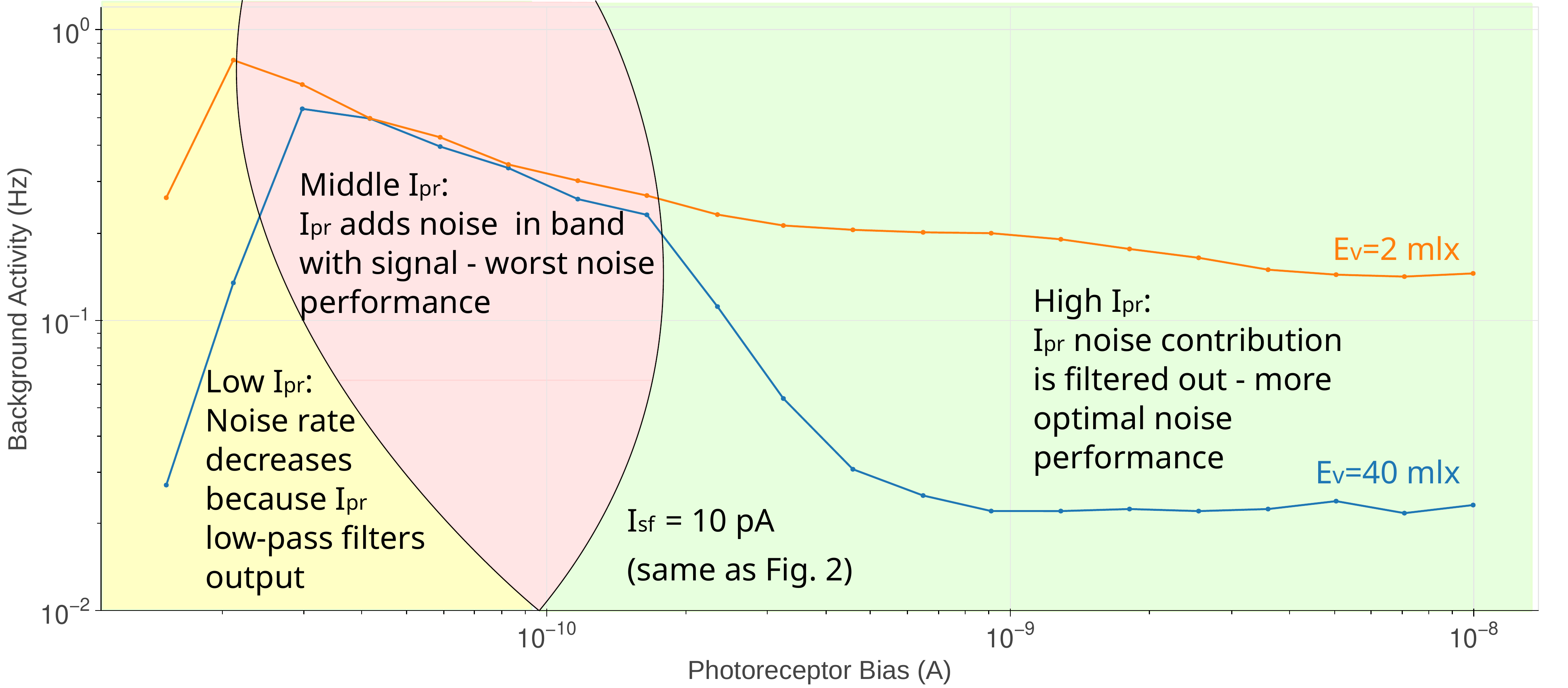}
\caption{Background activity measured from a DAVIS346 test pixel under constant on-chip illuminance of \SI{2}{\milli\lux} (orange line)  and \SI{40}{\milli\lux} (blue line) for $\isf$ of \SI{10}{\pico\ampere} (as in \cref{fig:psd}) and nominal threshold and refractory bias settings (see \cite{graca2023shininglight} for a characterization of these parameters, nominal settings corresponds to tweaks of 0 there).}
\label{fig:noise_rates}
\end{figure*}

\cref{fig:noise_rates} show the noise rates measured from the same test pixel for varying $\ipr$ for two different on-chip illumination levels. We observe that for high $\ipr$, noise rate becomes mostly constant, since all the noise components of $\ipr$ are filtered out. For middle $\ipr$ values, the noise contributions of $\ipr$ lie within the signal bandwidth and are not filtered out, and the noise rates peak. For lower $\ipr$ the noise rates decrease because $\ipr$ limits the bandwidth.

\subsection{Optimal biasing and optimality analysis}
From \cref{fig:psd}\textbf{c} we can see how strongly biasing $\ipr$ results in shifting the noise components added by $\ipr$ to higher frequencies outside the bandwidth of interest for signal. This means that we can filter them out using the \tip{SF} without consequences for signal. In the limit, if we bias $\ipr$ so strongly that all its contribution is removed by \tip{SF}, the output noise consists of only the noise contribution of $\ipd$ (which consists itself on equal parts of photon shot noise and $\Mfb$ noise), and the much smaller noise contribution of $\isf$. In this case, we are theoretically limited to a minimum of 2x photon shot noise when the contribution of $\isf$ becomes negligible.

The clear advantage of strongly biasing $\ipr$ is illustrated in \cref{fig:psd}\textbf{f}. For $\ipr$ of \SI{3}{\nano\ampere}, the model predicts a contribution of photon shot noise of 46\% (approximating the theoretical limit of 50\%), resulting in a noise event rate of \SI{0.02}{\hertz} under nominal threshold and refractory biases~\cite{graca2023shininglight} versus 12\% for $\ipr$ of \SI{10}{\pico\ampere}, resulting in a noise event rate of \SI{0.66}{\hertz}. 

The model predicts an \tip{RMS} noise contribution equivalent to TC log-e units of 0.006 for $\isf$. The contributions of $\ipd$ and $\ipr$ depend on filtering, but for the case where the pole controlled by $\ipd$ is dominant and filtering by the \tip{SF} is not considered, they are respectively 0.04 and 0.06 for most values of $\ipd$ and $\ipr$. Although \tip{RMS} noise alone is not enough to characterize \tip{DVS} noise, since it does not contain information about the noise frequency~\cite{graca2021unravelingtheparadox}, these numbers are useful to evaluate design limitations to the event sensitivity (i.e. the minimum event threshold with acceptable noise rates). One important conclusion is that $\isf$ should be adjusted to the minimum acceptable bandwidth for each application and $\ipr$ should be adjusted so that all its contributions are filtered out. Given that increasing $\ipr$ increases power consumption, $\ipr$ should be optimized to trade off power with noise performance. In the limit where the photoreceptor bandwidth is much higher than the \tip{SF} bandwidth (which happens for very high illuminance, high $\ipr$ and nominal or low $\isf$) the noise, noise introduced by $\ipd$ and $\ipr$ is filtered out and \tip{SF} becomes the main noise contributor.

Filtering with \tip{SF} and not with $\ipr$ is generally a better idea since it introduces significantly less noise, and the noise it introduces is not filtered out in any case. However, in practical \tip{DVS} implementations operating in very dark settings, very low $\ipr$ may result in a lower bandwidth than the minimum achievable by the \tip{SF}, and minimizing both $\ipr$ and $\isf$ may result in less \tip{ba}.

\section{Conclusion}
The measurements and analysis presented show that the \tip{DVS} pixel is limited to a minimum of 2x photon shot noise, and that using high $\ipr$ and adequate $\isf$ approximates this limit. We also discuss the limits imposed to event sensitivity by each noise contributor.


\renewcommand*{\bibfont}{\footnotesize}
\footnotesize{\printbibliography}

\end{document}